\documentclass[final]{svjour2}
\usepackage{graphicx}
\graphicspath{    
{imagesPoster/}
{images/}
} 
\usepackage{rotating}
\usepackage{amssymb}
\usepackage{mathptmx}
\usepackage[numbers]{natbib}
\makeatletter
\journalname{Journal of Low Temperature Physics}

\bibpunct{}{}{,}{s}{}{,}

\begin{document}

\newcommand{\urs}[1]{\ensuremath{_{\mathrm{#1}}}}
\newcommand{\ursb}[1]{\ensuremath{\mathrm{#1}}}

\title{A Highly Linear Calibration Metric for TES X-ray Microcalorimeters}

\author{C.G. Pappas, J.W. Fowler, D.A. Bennett, W.B. Doriese,  Y.I. Joe, K.M. Morgan, G.C. O'Neil, J.N. Ullom, D.S. Swetz}

\institute{Quantum Sensors Group, NIST Boulder Laboratories, 325 Broadway, MS 687.08, Boulder, Colorado 80305, USA. Tel: +1 303-497-3874.\\
\email{christine.pappas@nist.gov}}

\maketitle

\begin{abstract}

Transition-edge sensor X-ray microcalorimeters are usually calibrated empirically, as the most widely-used calibration metric, optimal filtered pulse height (OFPH), in general has an unknown dependance on photon energy, $E\urs{\gamma}$. Because the calibration function can only be measured at specific points where photons of a known energy can be produced, this unknown dependence of OFPH on $E\urs{\gamma}$ leads to calibration errors and the need for time-intensive calibration measurements and analysis. A calibration metric that is nearly linear as a function of $E\urs{\gamma}$ could help alleviate these problems. In this work, we assess the linearity of a physically motivated calibration metric, $E\urs{Joule}$. We measure calibration pulses in the range 4.5 keV$<$$E\urs{\gamma}$$<$9.6 keV with detectors optimized for 6 keV photons to compare the linearity properties of $E\urs{Joule}$ to OFPH. In these test data sets, we find that $E\urs{Joule}$ fits a linear function an order of magnitude better than OFPH. Furthermore, calibration functions using $E\urs{J}$, an optimized version of $E\urs{Joule}$, are linear within the 2-3 eV noise of the data.

\keywords{Microcalorimeter, Transition-edge sensor (TES), Detector calibration, X-ray spectroscopy}

\end{abstract}

\section{Introduction}

The transition-edge sensor (TES) microcalorimeter is capable of detecting single photons with energy resolving powers exceeding 1000. It can be optimized for use in the soft X-ray, hard X-ray, or gamma ray regime~\cite{XrayTESReview}.TES microcalorimeters have been used for a variety of applications~\cite{RSI}, including particle-induced X-ray emission (PIXE)~\cite{pixe}, time-resolved chemistry~\cite{tabletop1}~\cite{tabletop2}, nuclear forensics~\cite{nuc_foren}, and hadronic atom studies~\cite{hadron}. Other experiments in development include instruments for indirect measurement of the neutrino mass~\cite{holmes} and orbiting astronomical telescopes~\cite{Athena}~\cite{Athena_dets}~\cite{Lynx}.

In all these applications, the TES microcalorimeter requires a calibration procedure to map the directly measured signal, a time record of the pulse in the TES current, into the desired quantity: photon energy ($E\urs{\gamma}$). Various calibration metrics can be calculated from the TES current pulse as a function of $E\urs{\gamma}$ to generate a calibration function. The most commonly used metric is the optimal filtered pulse height (OFPH), a low-noise estimator of the pulse size~\cite{optfilter}~\cite{pulseprocess}.

When pulse signals are very small, the detector properties remain at their quiescent values during the pulse. In this small-signal limit, the pulse shape is constant with a size proportional to $E\urs{\gamma}$, and the OFPH is proportional to $E\urs{\gamma}$. However, a TES microcalorimeter can only be designed to remain in the small-signal limit over a large energy range at the sacrifice of energy resolution. 

In reality, the OFPH calibration function has a complicated shape that is determined completely empirically. It can only be measured at select points where photons can be generated with known energies. The calibration function must then be interpolated (and extrapolated) smoothly between (and beyond) these measured ``anchor points".  Even when there are anchor points close to the unknown energy and careful interpolation procedures are used, calibration done in this way can be the leading source of systematic uncertainty on absolute line energies~\cite{metrology}.

A metric with a known dependence on $E\urs{\gamma}$ (up to some fitting parameters) could produce more accurate calibrations, reduce requirements on anchor points, and simplify calibration analysis. This has led to research into calibration metrics that are more linear than OFPH as a function of $E\urs{\gamma}$~\cite{Rspace}~\cite{Rspace2}~\cite{Peille}~\cite{hollerith}. In this paper, we discuss a calibration metric, $E\urs{Joule}$, that is nearly proportional to $E\urs{\gamma}$ when applied to our hard X-ray detectors under typical operating conditions. We also discuss a slight adjustment to $E\urs{Joule}$ that produces the more linear $E\urs{J}$ metric. We concern ourselves only with the definition and linearity properties of $E\urs{Joule}$ and $E\urs{J}$. The direct application to individual, noisy pulse records of the formulas here will yield linear, but highly noisy estimations of $E\urs{\gamma}$. We treat the problem of their statistically optimal estimation in a companion paper in these same proceedings~\cite{low_noise_EJ}.

\section{The $E\urs{Joule}$ Metric}

A TES microcalorimeter measures the energy of individual photons. The photon is absorbed by an ``island" that is only weakly thermally connected to a silicon substrate held at constant temperature, $T\urs{bath}$. The signal resulting from an absorbed photon is a downward pulse in the current through the voltage-biased TES on the island. The shape of this signal depends on the thermal properties of the detector, the  electrical properties of the TES bias and readout circuits, and the resistive transition function of the TES with respect to temperature and current, $R(T,I)$. Here, we parameterize the $R(T,I)$ function by the TES normal resistance ($R_n$), the TES critical temperature ($T\urs{c}$), and the local logarithmic derivatives of $R(T,I)$ with respect to temperature and current: $\alpha(T,I)$ and $\beta(T,I)$.

To define the $E\urs{Joule}$ metric, which is also referred to as $E\urs{ETF}$~\cite{irwinhilton}~\cite{hollerith}, we start with the standard ``one-body" thermal model describing this system~\cite{irwinhilton}~\cite{2body}:

\begin{equation}
	P\urs{\gamma}=P\urs{bath}-P\urs{Joule}+C\frac{dT}{dt},
\label{diff_thermal}
\end{equation}
where $P\urs{Joule}$ is the electrical power generated by the current through the TES, $P\urs{\gamma}$ is the photon power deposited on the TES island, $C$ is the total heat capacity of the TES island, and $T$ is the TES temperature, and $P\urs{bath}$ is the thermal power flowing from the TES island to the bath. This quantity is usually modeled as a power-law:

\begin{equation}
	P\urs{bath}=\kappa(T^n-T\urs{bath}^n),  \hspace{4mm} \frac{dP\urs{bath}}{dT}=G(T)=\kappa n T^{n-1},
\label{Pbath}
\end{equation}
where $\kappa$ and $n$ are constants.

Integrating Eq.~\ref{diff_thermal} over the duration of the pulse (from $t\urs{0}$ to $t\urs{f}$) yields an expression for the photon energy:

\begin{equation}
	E\urs{\gamma}=E_{bath}+E_{Joule}=\int_{t\urs{0}}^{t\urs{f}}{\left(P\urs{bath}-P\urs{Joule}\right)dt}=\int_{t\urs{0}}^{t\urs{f}}{\left(\Delta P\urs{bath}-\Delta P\urs{Joule}\right)dt}.
	\label{eq_Etotal}
\end{equation}
The final term of Eq.~\ref{diff_thermal} integrates to zero because the temperature is the same before and after the pulse. When there are no photons incident on the detector, $P\urs{bath}$ is equal to $P\urs{Joule}$. Therefore, we can also write Eq.~\ref{eq_Etotal} as shown on the right, where $\Delta P\urs{bath}$ and $\Delta P\urs{Joule}$ are the deviations of $P\urs{bath}$ and $P\urs{Joule}$ from their common quiescent value. We define the two positive quantities, $E\urs{bath}$ and $E\urs{Joule}$, as the integrals of $\Delta P\urs{bath}$ and -$\Delta P\urs{Joule}$, respectively, over the course of the pulse.

Our TESs are nearly voltage-biased to provide negative electrothermal feedback by applying a constant bias current, $I\urs{bias}$, to the TES in parallel with a relatively small shunt resistor, $R\urs{sh}$. The TES current is coupled to a SQUID in a flux-locked loop, then amplified~\cite{TDM}. We directly measure changes in the feedback current necessary to keep the SQUID on its lockpoint, $I_{FB}$, which we scale by a proportionality constant, $M_r$, to yield changes in the TES current, $I$: $\delta I=M_r \delta I_{FB}$. Inductance is often added in series with the TES to slow the current signal for ease of readout. We refer to the total inductance in the bias circuit as $L$. For a TES in this bias configuration, $\Delta P\urs{Joule}$ is given by the following expression:

\begin{equation}
	\Delta P\urs{Joule}=R\urs{sh}\left(\delta I (I\urs{bias}-2I\urs{0}) -\delta I^2\right) - (I\urs{0}+\delta I)\frac{d(\delta I)}{dt} L,
	\label{dPjoule}
\end{equation}
where $I\urs{0}$ is the quiescent value of the TES current before the pulse and $\delta I=I-I\urs{0}$. Integrating over the course of the pulse gives the following expression for $E\urs{Joule}$:

\begin{equation}
	E\urs{Joule}=-R\urs{sh}\left[(I\urs{bias}-2I\urs{0})\int_{t\urs{0}}^{t\urs{f}}{\delta Idt}  -\int_{t\urs{0}}^{t\urs{f}}{(\delta I)^2dt}\right].
\label{eq_EJ}
\end{equation}
The inductance term in Eq.~\ref{dPjoule} integrates to zero because the energy stored in the inductor is the same before and after the pulse.

One could in principle determine the energy of a photon absorbed by the detector with no need for a calibration curve by calculating both $E\urs{bath}$ and $E\urs{Joule}$ from the pulse signal (Eq. 3). The $E\urs{bath}$ term may be calculated by measuring the thermal conductivity constants $\kappa$ and $n$ and calculating the temperature of the TES at every point during the pulse (Eq. 2). Although the detector is not in thermal equilibrium during the pulse, the TES temperature can be determined from its previously measured $R(T,I)$ transition function (assuming the function $T(R,I)$ is single-valued and can be measured over the entire transition region the pulse will traverse). In practice, we have found that using our standard detector characterization techniques, $E\urs{\gamma}$ can be estimated to about $\pm$ 5$\%$ with this method, while many experiments would require precision on the order of $\pm$ 0.1$\%$. We have not yet explored the fundamental limits of this method's precision, which may be constrained by how precisely it is possible to measure detector parameters and thermal ``two-body" effects arising from the finite thermal conductivity between the TES and absorber~\cite{2body}.
 
In contrast, calculation of the $E\urs{Joule}$ term up to a proportionality constant only requires knowledge of the quantities $I\urs{bias}$, $I\urs{0}$, and $M_r$, which in principle can all be determined to high precision with simple measurements. Although $E\urs{Joule}$ only approaches $E\urs{\gamma}$ in the infinite loop gain ($\mathcal{L}=\frac{P\urs{Joule}\alpha}{GT}$), small signal limit, we find under many conditions that $E\urs{Joule}$ is nearly proportional to $E\urs{\gamma}$. In these cases, $E\urs{Joule}$ is useful as a highly linear calibration metric. 

\section{Linearity of the $E\urs{Joule}$ Metric: Simulations}

\begin{figure}[htbp]
\includegraphics[width=1.\linewidth, keepaspectratio]{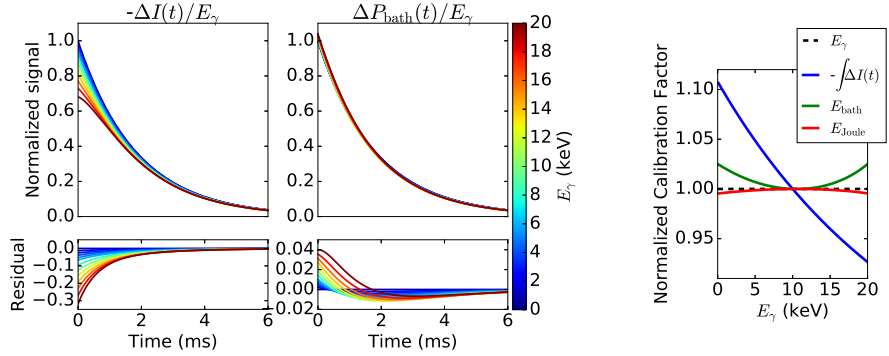} 
\caption{{\it Left:} Simulated pulses in $\Delta I/E\urs{\gamma}$ and $\Delta P\urs{bath}/E\urs{\gamma}$ space using the model described in the text. Residuals are taken with respect to the 1 keV photon signal. {\it Right:} The calibration factor is the metric divided by $E\urs{\gamma}$, then scaled to equal 1.0 at 10.0 keV. (Color figure online)}
\label{fig2fluid}
\end{figure}

When signals are large enough that OFPH is no longer a linear calibration metric, the non-linearized electrical and thermal differential equations describing the detector system must be solved to describe the response of a TES microcalorimeter to an absorbed photon~\cite{irwinhilton}. Because the solutions are difficult to describe analytically, in this section we explore the linearity of the $E\urs{Joule}$ metric through simulations of our typical TES hard X-ray microcalorimeters. For intuition-building purposes, we will consider the curvature of $E\urs{bath}(E\urs{\gamma})$ instead of $E\urs{Joule}(E\urs{\gamma})$. Because we operate under high loop gain conditions, it is almost always true that $E\urs{bath}<E\urs{Joule}$. When this is the case, by Eq.~\ref{eq_Etotal}, the curvature of the $E\urs{Joule}(E\urs{\gamma})$ calibration curve is less than or equal to the curvature of $E\urs{bath}(E\urs{\gamma})$, where we define the curvature of $F(E\urs{\gamma})$ as $\frac{d^2F}{dE_{\gamma}^2}\left(\frac{E_{\gamma}}{F}\right)^2$.

In our simulation, we use detector parameters of the ar13-9b devices~\cite{RSI} that produced the data in Section 4 of this paper when possible: $\kappa$=4000 pW/$\ursb{K^n}$, $n$=3.3, $R\urs{sh}$ =360 $\ursb{\mu}$$\ursb{\Omega}$, $T\urs{c}$=109 mK, $R\urs{n}$=12 m$\ursb{\Omega}$, and $T\urs{bath}$=65 mK. We describe the TES $R(T,I)$ function using the Two-Fluid model~\cite{2fluid}, with parameters that do not reproduce the $R(T,I)$ transitions of the ar13-9b TESs exactly, but yield $\alpha$ and $\beta$ values typical of these devices in the relevant transition area: $c\urs{I}*I_{c0}$=5.0 mA and $c\urs{R}$=0.5. Because $L$ does not affect $E\urs{bath}$ or $E\urs{Joule}$ by Eq.~\ref{eq_Etotal} and Eq.~\ref{eq_EJ}, we choose $L=0$ for convenience. We approximate the TES heat capacity as a constant value of 0.85 pJ/K, independent of temperature. Simulations using more precise measurements of our detectors' $C(T)$ functions may be included in future publications.

In this model, the TES temperature rises instantaneously by an amount $E\urs{\gamma}/C$ after a photon is absorbed, then decreases back to the quiescent temperature. Therefore, the height of the TES temperature pulse height is proportional to $E\urs{\gamma}$. The linear approximation of the $P\urs{bath}$ pulse height is $G\Delta T$, which is also proportional to $E\urs{\gamma}$. The actual $P\urs{bath}$ pulse height is very close to this linear approximation, such that it is linear with respect to photon energy within a few percent up to 20 keV (Fig.~\ref{fig2fluid}). In contrast, the $\Delta I$ pulse height differs from a linear relationship with $E\urs{\gamma}$ quite strongly. As shown in Fig.~\ref{fig2fluid}, this makes the $E\urs{bath}$ calibration function much closer to linear than that of the pulse average, $\int{-\Delta I(t)}dt$. The curvature of $E\urs{Joule}(E\urs{\gamma})$ is even smaller than $E\urs{bath}(E\urs{\gamma})$, because $E\urs{Joule}$ is about five times larger than $E\urs{bath}$.

\section{Linearity of the $E\urs{Joule}$ Metric: Experimental Results}

\begin{figure}[htbp]
\begin{center}
\includegraphics[width=.7\linewidth, keepaspectratio]{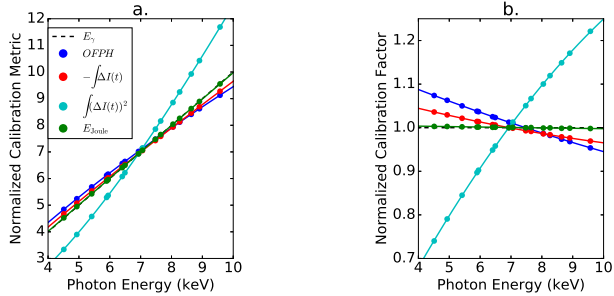} 
\end{center}
\caption{Example calibration data set from one of the ar13-9b detectors biased at 10$\%$ $R_n$ with $T\urs{bath}$=65 mK. Lines are 3rd order polynomial
fits to the data to guide the eye. Fig2a: Various calibration metrics as a function of $E\urs{\gamma}$, normalized as discussed in the text. Fig2b: Data from Fig2a is divided by $E\urs{\gamma}$ to yield the Calibration Factor. (Color figure online.)}
\label{figEJex}
\end{figure}

We have tested the linearity of the $E\urs{Joule}$  and comparison metrics on calibration data taken with 13 nominally identical ar13-9b X-ray microcalorimeters~\cite{RSI}. These detectors have the same properties used in the simulations above except for the $L$ values, which range from about 90 nH to 500 nH. Calibration data were taken with the detectors operated at a bath temperature of 65~mK at three different bias points, generating 39 data sets. Each calibration data set consists of averaged pulse signals from photons at the K$\alpha$ and K$\beta$ emission lines of the elements Ti, Cr, Mn, Fe, Co, Ni, Cu, and Zn, which are in the energy range of about 4.5 to 9.6 keV. 

A calibration function is produced by applying one of the metrics to a calibration data set. To quantify the linearity of each calibration function in units of eV, we compute a quantity we call linearity-$\sigma$ as follows. First, we fit the slope of a linear function intersecting the origin to the raw (metric vs. $E\urs{\gamma}$) calibration function. Then, we divide the metric values by this slope to obtain a normalized calibration function like the one in Fig.~\ref{figEJex}a. Finally, we compute the root-mean-square deviation of the normalized data set with respect to the $y=x$ line. 

\begin{figure}[htbp]
\includegraphics[width=1.\linewidth, keepaspectratio]{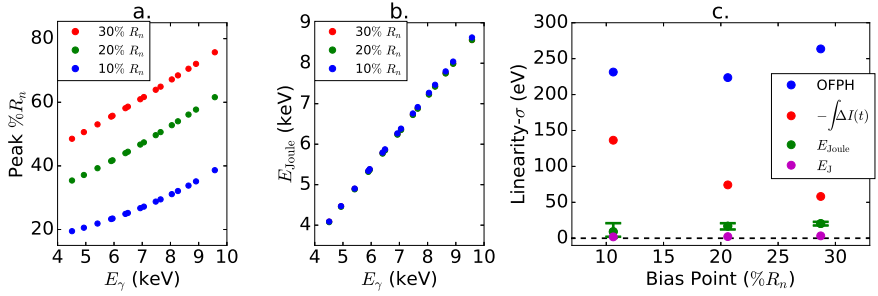} 
\caption{Calibration data taken with ar13-9b detectors at $T\urs{bath}$=65 mK and three different bias points. Fig3a: Data from one of the detectors with $L$ of about 500 nH. For each of the bias points, the TES resistance value at the peak of each pulse is plotted vs. $E\urs{\gamma}$. Fig3b: Using the same data as in Fig3a, $E\urs{Joule}$ is calculated and plotted vs. $E\urs{\gamma}$. Fig3c: For each metric, the linearity-$\sigma$, defined in the text, is averaged over the 13 ar13-9b detectors at each bias current, and plotted vs. the average bias $\% R_n$. The systematic error for $E\urs{Joule}$ is given by the error bars. It is calculated assuming the following measurement errors: $I\urs{bias}$: $\pm$3$\%$, $I\urs{0}$: $\pm$5$\%$, $M\urs{r}$: $\pm$3$\%$. The systematic error in the linearity-$\sigma$ of the other metrics is zero. The linearity-$\sigma$ values for the $\int{(\Delta I)^2}$ metric, not shown, are larger than 1 keV.  (Color figure online.)}
\label{figEJ}
\end{figure}

As shown in Fig.~\ref{figEJ}c, the $E\urs{Joule}$  metric generates calibrations functions with a linearity-$\sigma$ of 25 eV or less at all bias points tested, an order of magnitude better than OFPH. The $E\urs{Joule}$ metric is a linear combination of the pulse average and the pulse RMS, $\int{(\Delta I)^2dt}$. The pulse average tends to have negative curvature, while the pulse RMS tends to have positive curvature, as shown in Fig.~\ref{figEJex}. We define the metric $E\urs{J}$ as the linear combination of pulse average and pulse RMS that minimizes the linearity-$\sigma$ of the calibration function: $E\urs{J}= A\int{-\Delta I dt}+B\int{\Delta I^2 dt}$. The ratio $A/B$ is calculated separately for each detector and bias point. As shown in Fig.~\ref{figEJ}c, the linearity-$\sigma$ of the $E\urs{J}$ metric is only about 2-3 eV. This is within the noise of this measurement and two orders of magnitude better than that of the OFPH metric.

\section{Conclusions and Future Work}

We have explored the linearity of two calibration metrics, $E\urs{Joule}$ and $E\urs{J}$, when applied to hard X-ray TES microcalorimeters under typical operating conditions. On test data, we have found that the $E\urs{Joule}(E_{\gamma})$ calibration functions evaluated over ~4.5 keV to 9.6 keV fit a linear function an order of magnitude better than OFPH($E_{\gamma}$). An easy correction to $E\urs{Joule}$ yields a metric, $E\urs{J}$, that is linear within the 2-3 eV noise of the data. 

Because the $E\urs{J}$ metric produces nearly linear calibration functions, it may simplify calibration measurements and analysis and produce more accurate results. In a companion paper in this issue, a low-noise estimator of the $E\urs{J}$ metric is presented that produces Mn, Co, and Cu K$\alpha$ spectra with energy resolutions comparable to the OFPH results on test data~\cite{low_noise_EJ}. Our next step will be to compare the accuracy and speed of the full calibration procedures using OFPH, the low-noise $E\urs{J}$ estimator, and other proposed alternative calibration metrics such as resistance-space~\cite{Rspace}~\cite{Rspace2}~\cite{Peille}. 

The $E\urs{J}$ calibration method could potentially be used for a wide range of TES microcalorimeter applications. As the computational power required for the low-noise $E\urs{J}$ estimator is comparable to OFPH, these applications could include X-ray satellite missions. The $E\urs{J}$ metric may also be used to develop faster and more robust calibration analysis algorithms. Currently, due to the complexity of the procedure, calibration of data from TES spectrometers is typically performed after the completion of an experiment. The $E\urs{J}$ metric could be applied to software in development that will give TES spectrometer users calibrated results in real time to inform their next steps.

\begin{acknowledgements}
This work was supported by NIST's Innovations in Measurement Science Program and NASA SAT NNG16PT181. C.G.P is supported by the National Research Council Post-Doctoral Fellowship. As this is a contribution of a U.S. Government agency, it is not subject to copyright in the USA.
\end{acknowledgements}

\end{document}